\documentclass[11pt]{elsart}
\def \irbaddress{Rudjer Bo\v{s}kovi\'{c} Institute, Bijeni\v{c}ka cesta 54, P.O. Box 180, 10002 Zagreb, Croatia}
\def \untzaddress{University of Tuzla, Faculty of Science, Univerzitetska 4, 75000 Tuzla, Bosnia and Herzegovina}
\def \mainzaddress{Institut f\"{u}r Kernphyik, Universit\"{a}t Mainz, D-55099 Mainz, Germany}
\def \GWUSAIDaddress{Data Analysis Center at the Institute for Nuclear Studies, Department of Physics, The George Washington University, Washington, D.C. 20052}
\usepackage[usenames]{color}
\usepackage{epsfig}
\usepackage{amssymb,amsmath}
\usepackage{subfig}

\definecolor{grey}{rgb}{0.9,0.9,0.9}
\definecolor{black}{rgb}{0,0,0}

\begin{document}

\begin{frontmatter}



\title{ Generalization of the model-independent Laurent-Pietarinen single-channel pole-extraction formalism to multiple channels}


\author[IRB]{A. \v Svarc\corauthref{ca}},
\corauth[ca]{Corresponding author}
\ead{svarc@irb.hr}
\author[UNTZ]{Mirza Had\v{z}imehmedovi\'{c}},
\author[UNTZ]{Hedim Osmanovi\'{c}},
\author[UNTZ]{Jugoslav Stahov},
\author[MAINZ]{Lothar Tiator},
\author[GWUSAID]{Ron L. Workman},
\address[IRB]{\irbaddress}
\address[UNTZ]{\untzaddress}
\address[MAINZ]{\mainzaddress}
\address[GWUSAID]{\GWUSAIDaddress}

\begin{abstract}
A method to extract resonance pole information from single-channel partial-wave amplitudes  based on a Laurent (Mittag-Leffler) expansion and conformal mapping techniques has recently been developed. This method has been applied to a number of reactions and provides a model-independent extraction procedure
which is particularly useful in cases where a set of amplitudes is available only at descrete energies. This method has been generalized and applied to
the case of a multi-channel fit, where several sets of amplitudes are analysed simultaneously. The importance of unitarity constraints is discussed. The
final result provides a powerful, model-independent tool for analyzing partial-wave amplitudes of coupled or connected channels based entirely on
the concepts of analyticity and unitarity.
\end{abstract}
\end{frontmatter}

\newpage
The Particle Data Group (PDG)  \cite{RPP2014} has begun to include and emphasize the importance of pole-related quantities, de-emphasizing and eliminating many Breit-Wigner parameters, as the link between experiment and QCD. As a result, the analytic structure of theoretical and experimental partial-wave amplitudes in the complex energy plane has become increasingly important. A common approach involves the construction and solution of elaborate theoretical models, with free parameters fitted to available sets of experimental data. These can then be analytically continued into the complex energy plane.  Attempts to evade the  ``single-user" drawback of such approaches\footnote{A typical model is, due to its complexity, usually solvable and verifiable by one group only.} have involved simpler single-channel  pole extraction methods such as the speed plot \cite{Hoehler93}, time delay \cite{Kelkar}, the N/D method \cite{ChewMandelstam}, regularization procedures \cite{Ceci2008}, and Pade approximants \cite{Padde}. However, success has been limited.  As a step forward, a simple but quite reliable, model-independent single-channel pole-extraction formalism has been constructed, based entirely on principles of analyticity and unitarity. This method was named the Laurent+Pietarinen (L+P) expansion \cite{L+P2013}, and is based on an early application of these principles in the analysis of pion-nucleon scattering data \cite{Ciulli,CiulliFisher, Pietarinen,Pietarinen1}.
\\ \\ \indent
In spite of the fact that this single-channel L+P method is now generally applicable, extensively used in a wide array of problems  \cite{L+P2014,L+P2015,L+P2014a}, and already recognized by PDG as a confident tool for extracting pole positions of most baryon resonances \cite{PDG2015}, all applications in which one pole couples to several correlated quantities are still beyond its reach. For example, correlated multipoles in $\pi$ and $\eta$ photoproduction, and  partial wave amplitudes in coupled-channel models can only be treated in a sequence of independent single-channel procedures, missing the constraint that poles in all such situations must be the same. Also, in some cases, all existing poles may not be recognized in each individual process, and that in particular happens if a resonance coupling to a particular channel is weak.  Thus, the main purpose of this paper is to create a new method which enables the treatment of all connected channels simultaneously. We have generalized the existing single-channel L+P formalism (SC L+P) to the multi-channel case (MC L+P) in such a way that pole positions are unique, but with differing residua which are to be related to branching fractions. This also allows the analysis of photo- and electro-production in which a single pole contributes to two or three multipoles.  Just as in the single-channel L+P method, the most important application of the method would be the analysis of partial wave data (discreet quantities obtained directly from experiment, with very few stabilizing theoretical assumptions), rather than treating the partial wave amplitudes which are coming from theoretical calculations. Therefore, this method as such represents the first model-independent way to treat multi-channel experimental data directly, and is extremely important for precise and rapid analysis of new ongoing experimental programs.
\\ \\ \indent
The driving concept behind the single-channel L+P approach was to replace solving an elaborate theoretical model and analytically
continuing its solution  into the full complex energy plane, with a local power-series representation of partial wave amplitudes given on the real energy axis. In such a way, the global complexity of a model is replaced by much simpler model-independent expansion limited to the regions near the real energy axis which is sufficient to obtain poles and their residues. Formally, the introduced L+P method was based on the Mittag-Leffler expansion\footnote{Mittag-Leffler expansion \cite{Mittag-Leffler} is the generalization of a Laurent expansion to a more-than-one pole situation. For simplicity, we will simply refer to this as a Laurent expansion.} of partial wave amplitudes near the real energy axis, representing the regular, but unknown, background term by a conformal-mapping-generated, rapidly converging power series called a Pietarinen expansion\footnote{A conformal mapping expansion of this particular type was introduced by Ciulli and Fisher \cite{Ciulli,CiulliFisher}, was described in detail and used in pion-nucleon scattering by Esco Pietarinen \cite{Pietarinen,Pietarinen1}. The procedure was denoted as a Pietarinen expansion by G. H\"{o}hler in \cite{Hohler84}.}. In practice we have represented the regular background part with three Pietarinen expansion series, and fitted all free parameters in our approach to the chosen channel input. The first Pietarinen expansion with branch-point $x_P$ was restricted to an unphysical energy range and represented all left-hand cut contributions, and next two Pietarinen expansions described background in the physical range with branch-points $x_Q$ and $x_R$ defined by the analytic properties of the analyzed partial wave. A second branch-point was usually fixed to the elastic channel branch-point, and the third one was either fixed to the dominant channel threshold value or left free.  Thus, solely on the basis of general physical assumptions about analytic properties of the fitted process (number of poles and number and position of conformal mapping branch-points) the pole parameters in the complex energy plane are obtained. In such a way, the simplest analytic function with a set of poles and branch-points which is fitting the input is actually constructed. This method is equally applicable to both theoretical and experimental input\footnote{Observe that fitting partial wave data coming from experiment is even more favorable.}, and represents the first reliable procedure to extract pole positions from experimental data, with minimal model bias.
\\ \\ \indent
The generalization of L+P method to MC L+P is performed in the following way: i) we have made separate Laurent expansions for each channel (coupled quantity); ii) we have kept pole positions fixed for all channels (quantities), iii) we have left all residua and all Pietarinen coefficients free; iv) we have chosen the branch-points exactly as we would for the single-channel model; v) we have generalized the single-channel discrepancy function $D_{dp}^a$ (see Eq. (5) in ref. \cite{L+P2015}) which quantifies the deviation of the fitted function from employed input to a multi-channel quantity $D_{dp}$ by summing up all single-channel contributions, and vi) the minimization is performed for all channels of the input in order to obtain the final solution.
\\ \\ \noindent
The final model can be summarized by the following set of formulae for $k$ resonances:

\begin{eqnarray}
\label{eq:Laurent-Pietarinen}
T^a(W) &=& \sum _{i=1}^{k} \frac{x^{a}_{i} + \imath \, \, y^{a}_{i}  }{W-W_i} + \nonumber \\
       & + & \sum _{l=0}^{L^a}  c^a_l \, X^a (W )^l  +  \sum _{m=0}^{M^a} d_m^a \, Y^a (W )^m +  \sum
_{n=0}^{N^a} e_n^a \, Z^a (W )^n \nonumber \\
X^a (W )&=& \frac{\alpha^a-\sqrt{x_P^a-W}}{\alpha^a+\sqrt{x_P^a - W }}; \, \, \, \, \,   Y^a(W ) =  \frac{\beta^a-\sqrt{x_Q^a-W }}{\beta^a+\sqrt{x_Q^a-W }};  \, \, \, \, \,
Z^a(W ) =  \frac{\gamma^a-\sqrt{x_R^a-W}}{\gamma^a+\sqrt{x_R^a-W }}
 \nonumber \\
 D_{dp} & = & \sum _{a}^{all}D_{dp}^a \nonumber \\
 D_{dp}^a & = & \frac{1}{2 \, N_{data}} \, \, \sum_{i=1}^{N_{data}} \left\{ \left[ \frac{{\rm Re} \, T^{a}(W_i)-{\rm Re} \, T^a_{exp}(W_i)}{ Err_{i,a}^{\rm Re}}  \right]^2 + \right. \nonumber \\
 &+&    \left. \left[ \frac{{\rm Im} \, T_{a}(W_i)-{\rm Im} \, T^a_{exp}(W_i)}{ Err_{i,a}^{\rm Im}} \right]^2 \right\}  + {\cal P}^a \nonumber \\
 && {\cal P}^a \, \, .....   \, \, {\rm Pietarinen \, \, and \, \, unitarity \, \, penalty \, \, functions} \nonumber \\
 && Err_{i,a}^{\rm Re, \, Im} ..... {\rm \, \, minimization \, \, error\, \, of \, \, real \,\, and \, \, imaginary \, \, part \, \, respectively,} \nonumber \\
 && a \, \, .....   \, \,    {\rm  correlated \, \, quantity \, \, index}{ \, \, (\pi N \rightarrow \pi N , \, \pi N \rightarrow  \eta N,  E_{l_\pm}, M_{l_\pm}  ...)} \nonumber
\nonumber \\
&&  L^a, \, M^a, \, N^a \, ... \, \in  \mathbb{N} \, \, \, {\rm number \, \, of \, \, Pietarinen \, \, coefficients \, \, in \, \, channel \, \, \mathit{a} }.
 \nonumber \\
&&  W_i,\, W   \in   \mathbb{C}
\nonumber \\
&& x_i^a, \, y_i^a, \, c_l^a, \, d_m^a, \, e_n^a, \, \alpha^a, \, \beta^a, \, \gamma^a ... \in  \mathbb{R}  \, \,
 \nonumber \\
\nonumber
\end{eqnarray}

In the MC L+P formalism, the unitarity part of the single-channel penalty function ${\cal P}^a$ is also generalized. Namely, in the SC L+P formalism,
the penalty function $\cal P$ had two contributions: Pietarinen penalty function
\begin{center}
$\Lambda_{Piet} \, \cdot \, \sum_{j=1}^{3}\lambda^j \, \left( \sum _{k=1}^{N} (c_k^j)^2 \, k^3 \right)$
\end{center}
 and unitarity constraint in the physical region
 \begin{center}
 $\Lambda_{phys} \, \cdot \, \sum_{j=1}^{N_{pts}^{el}}(1-S(W_j)S(W_j)^\dagger)^2, $ 
  \end{center}
  \newpage
In the MC L+P extension, we have added two more terms to the penalty function: \\ \noindent
1. subthreshold unitarity for the elastic channel
\begin{center}
Im $T^{el}(W)$ \, = \, 0 for $W \, < W_{el-thr} $,  and
\end {center}
 2. subthreshold unitarity for other channels
\begin{center}
Im $T^{inel}(W)$ \, = \, 0 for $W \, < W_{inel-thr}$.
\end{center}

Observe that both additional conditions can be used only if our input quantities (partial waves or multipoles) are also unitarized. Adding subthreshold unitarity, when possible, improved our model significantly, and introducing it via penalty function enabled us to consider the importance of all unitarity restrictions explicitly by varying penalty function strength parameters $\Lambda$.
\\ \\  \noindent
\emph{Comparing Single- and Multi-channel Fits} \vspace*{0.2cm} \\
 \indent
 We have tested the validity of MC L+P model, and analyzed which new insights can be gained in a comparison of the SC L+P and MC L+P analyses by  applying it to  Bonn-Gatchina  BG 2011-2 P$_{11}$ $\pi N$ elastic and $\pi N \rightarrow \eta N$ amplitudes for which resonance pole parameters are provided through the analytic continuation of the Bonn-Gatchina model and published in \cite{BG2012,BG2012a}. We first show the problems which occur when two independent single-channel L+P analyses are performed on $\pi N$ elastic and $\pi N \rightarrow \eta N$ amplitudes, then demonstrate how the MC L+P approach solves these problems. Finally, we confirm the validity and precision of our method by comparing our results with known and published pole parameter values.  We do not expect to exactly reproduce the published pole results, as both the original SC L+P and new MC L+P methods are based on approximations to the analytic structure of the true non-resonant background functions. However, we do expect good agreement within the uncertainties of the various extraction methods.
\\ \\ \indent
We have first made two independent SC L+P analyses of $\pi N$ elastic and $\pi N \rightarrow \eta N$ amplitudes.  Results are shown in Fig.1 and Table \ref{SC L+P}.

\begin{center}
\begin{figure}[!h]
\includegraphics[width=6.5cm]{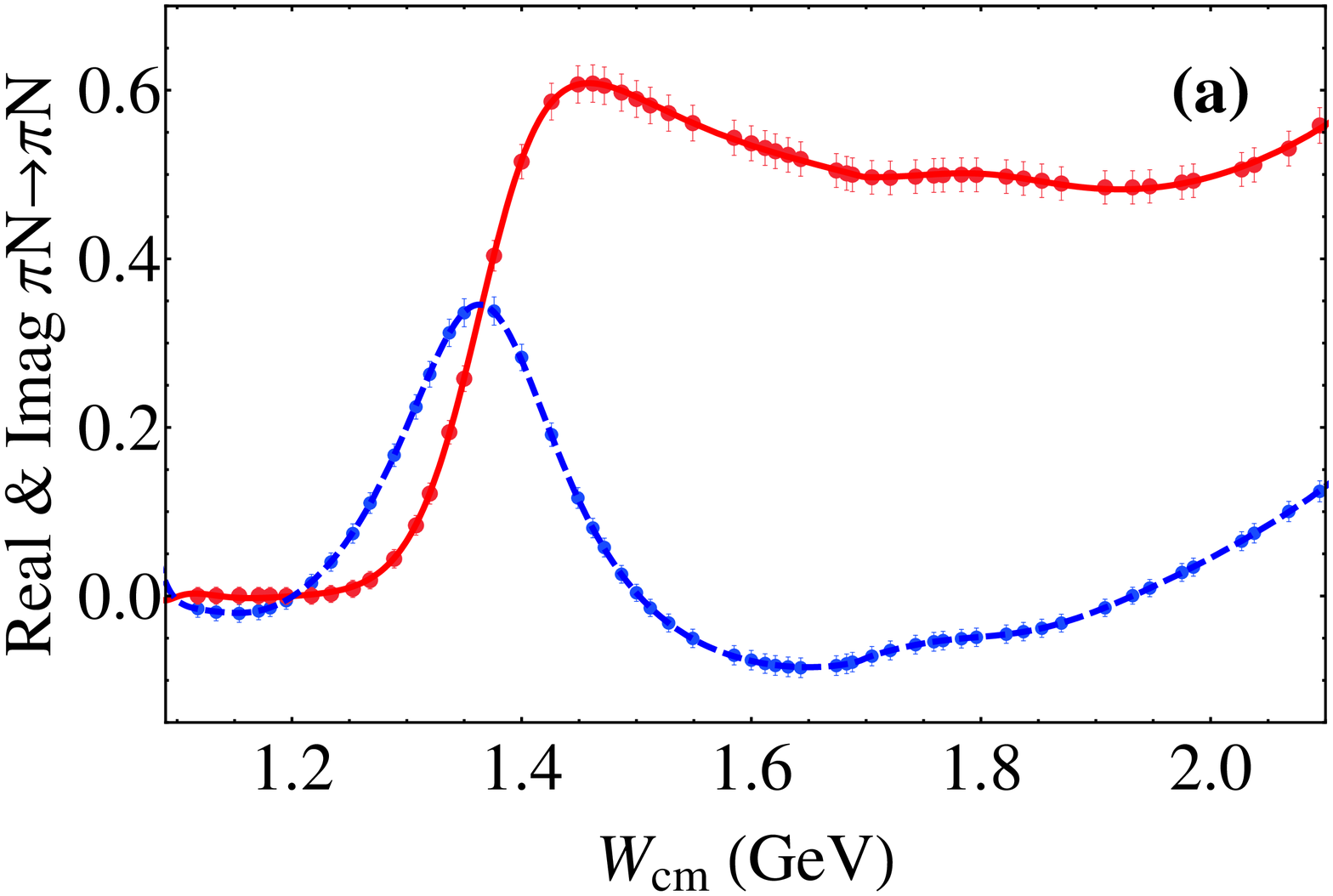} \hspace*{0.5cm}
\includegraphics[width=6.5cm]{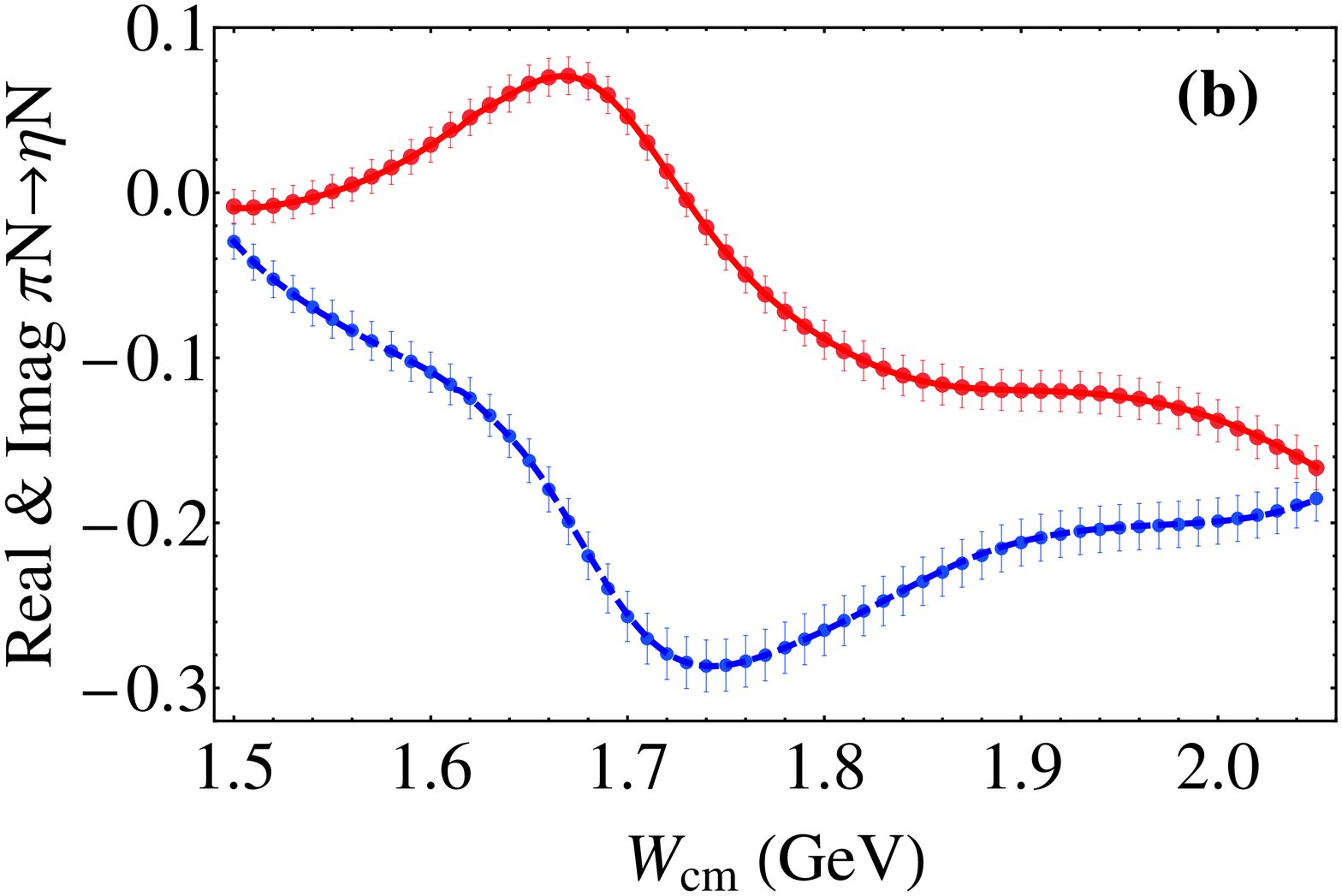}
\caption{\small(Color online) The SC L+P result for BG2011-2 \cite{BG2012,BG2012a}  $\pi N \rightarrow \pi N$ and $\pi N \rightarrow \eta N$  PW amplitudes is shown in (a) and (b) respectively. Blue and red full and dashed lines give the real and imaginary parts \vspace*{1.cm} respectively. }
\label{Fig1}
\end{figure}
\end{center}
\newpage
Here we see that the SC L+P fit for both reactions is very good with two poles only, but these poles are not identical.  The $\pi N$ elastic reaction can be fitted with the N(1440)1/2+ and N(1880)1/2+ while $\pi N \rightarrow \eta N$ can be fitted with the N(1710)1/2+ and N(1880)1/2+. In SC L+P fits there is no indication that the N(1710)1/2+ is needed in $\pi N$ elastic scattering, nor the sub-threshold N(1440)1/2+ in $\pi N \rightarrow \eta N$. Further, the presence of a fourth N(2100)1/2+ resonance is not indicated in either channel. Finally, numerical values for the second obtained resonance N(1880)1/2+ are different for each reaction. 
\\ 
\begin{table}[h!]
\caption{Two independent SC L+P analyses of $\pi N$ elastic and $\pi N \rightarrow \eta N$ BG 2011-2 amplitudes. \\ }
\label{SC L+P}
\begin{center}
 \begin{tabular}{c|c|ccccc}
\hline \hline
 \hspace*{0.2cm} Fitted  \hspace*{0.2cm}& Resonance     &  $M_i $  &  $\Gamma _i$  &   $|a_i|$ & $\theta$ & $D_{dp}^a$  \tabularnewline
\hspace*{0.2cm} channel \hspace*{0.2cm} &  name         &          &                &           &          &             \tabularnewline
 \hline \hline
$\pi N$ elastic                         &  N(1440)1/2+  &  1368    &     193       &     49    &     -82  &    0.004     \tabularnewline
 \cline{2-6}
      two poles                         &  N(1880)1/2+  &  1857    &     321       &      15   &     179  &               \tabularnewline
\hline
$\pi N \rightarrow \eta N$              &  N(1710)1/2+  &  1686    &     204       &     19    &     -27  &   0.002      \tabularnewline
 \cline{2-6}
      two poles                         &  N(1880)1/2+  &  1861    &     252       &      20   &     -95   &             \tabularnewline
\hline \hline
\end{tabular}
\end{center}
\end{table}

\begin{figure}[h!]
\includegraphics[width=6.5cm]{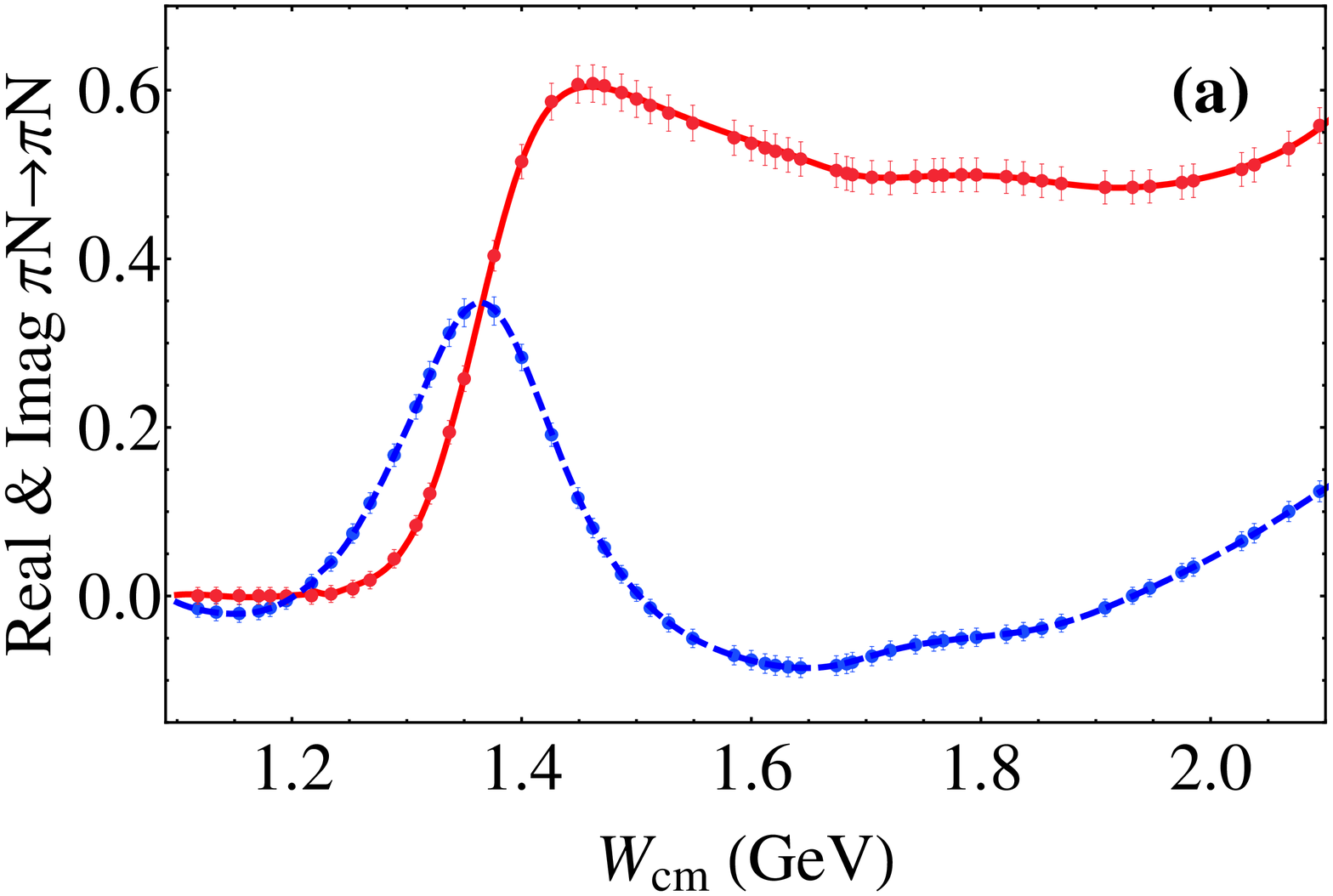} \hspace*{0.5cm}
\includegraphics[width=6.5cm]{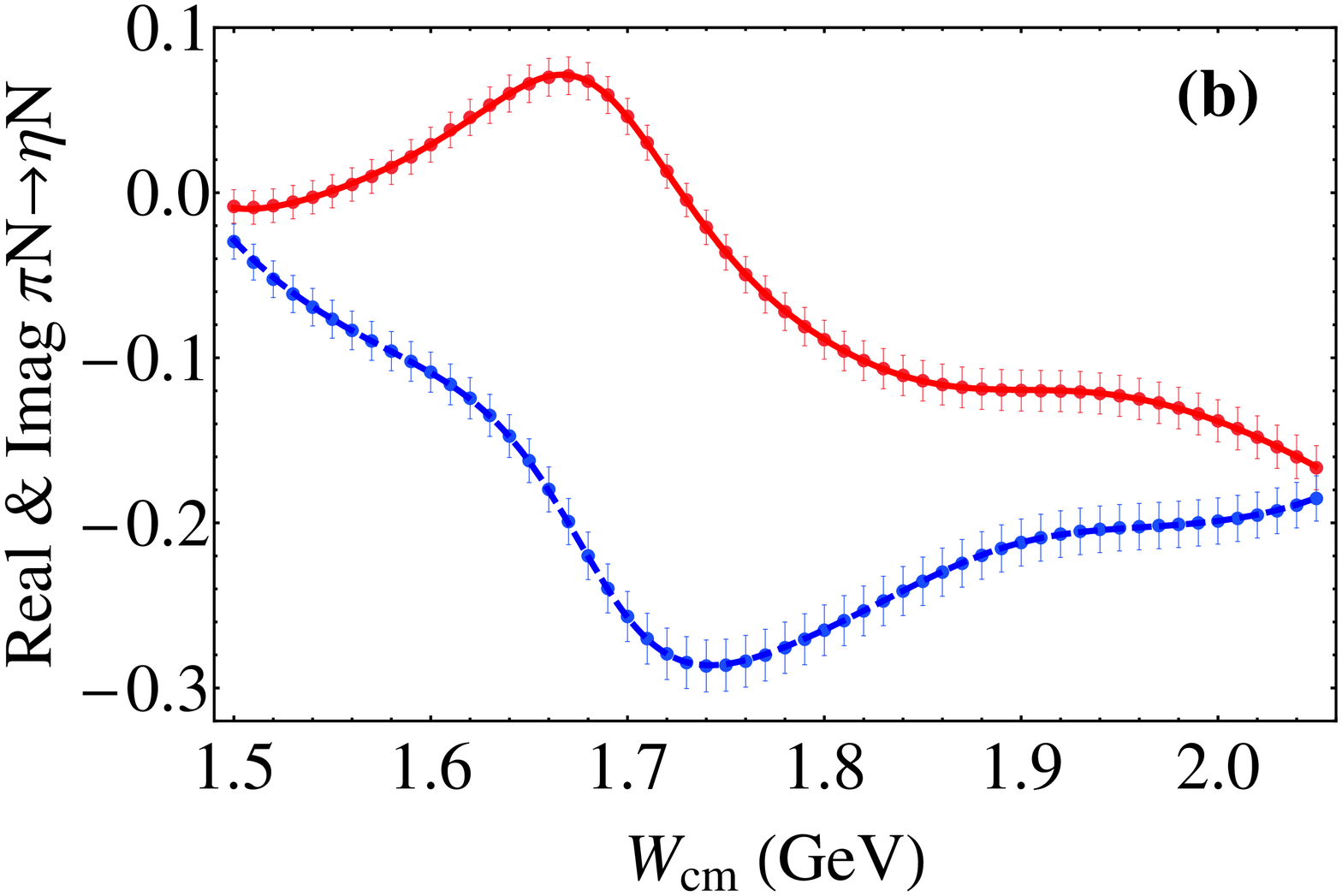} 
\caption{\small(Color online) The MC L+P result for BG2011-2 \cite{BG2012,BG2012a}  $\pi N \rightarrow \pi N$ and $\pi N \rightarrow \eta N$  PW amplitudes is shown in (a) and (b) respectively. Blue and red full and dashed lines give the real and imaginary parts \vspace*{1.cm} respectively. }
\label{Fig2}
\end{figure}

In a coupled-channel analysis this is not permissible. If a resonance exists in a certain partial wave in one reaction, it should exist in all reactions which couple to this partial wave. Therefore, both reactions should be fitted with at least three resonances with
consistent pole positions. This is achieved in the proposed MC L+P approach.
\\ \\ \\  \indent
In preliminary MC L+P fits to the BG 2011-2 amplitudes  \cite{BG2012}, we began by using three poles. This corresponded to the number of poles reported in reference \cite{BG2012}. To our surprise, the fit failed to obtain a good result with three resonances only. A good fit required the existence of a fourth state.  This puzzling result was understood once a careful scan of the literature revealed that the Bonn-Gatchina group indeed does consider a further pole  \mbox{(Re(pole), -2Imag(pole))} at (2100, 500) MeV which slightly improves the stability of the fit in their model \cite{BG2012a}. However, the new pole is poorly determined, and they neither claim its existence nor rule it out.
In effect, the MC L+P fit not only included all expected resonances in fitting both reactions, it predicted a fourth state as well which finally turned out to be allowed in the original BG 2011-2 model.
\\ \\ \noindent
Final results of MC L+P fit are shown in Fig. \ref{Fig2} and Table. \ref{MC L+P comparison}.

\begin{table}[h!]
 \caption{Comparison of published theoretical BG2011-2 \cite{BG2012,BG2012a} pole parameters with MC L+P  results. \\ }
\label{MC L+P comparison}
\begin{center}
\begin{tabular}{c|c|c|cc}
\hline \hline
       Resonance                    &                                   & PDG $^{[1]}$    &    BG$^{[18,19]}$       &  BG $^{\rm MC \, L+P}$     \tabularnewline
	      name                      &                                   &                 &                         &                            \tabularnewline \hline	
                                    &	$M$	                            &   1350-1380     &  1370(4) $^{[18]}$      &      1368 (3)               \tabularnewline
                                    &   $\Gamma$                        &     160-220     &   190(7)                &         191 (3)             \tabularnewline
                                    &  $ |a|_{\pi N}$	                &       40-52     &    48(3)                &          49 (2)           \tabularnewline
N(1440)1/2+                         &  $\Theta_{\pi N}$                 &       75-100    &    -78(4)               &          -82 (3)             \tabularnewline
\rule{0pt}{3ex}
                                    &  $\frac{2 |a|_{\eta N}}{\Gamma}$	&      -        &         -               &          0.1(0.1) \%                \tabularnewline
                                    &   $\Theta_{\eta N}$ 	            &        -        &          -              &            22(20)               \tabularnewline
\hline
	                                &       $M$	                        &   1670-1770     &     1687(17) $^{[18]}$  &       1686 (8)              \tabularnewline
                                    &   $\Gamma$                        &      80-330     &   200(25)               &          153 (24)             \tabularnewline
                                    &    $ |a|_{\pi N}$	                &       6-15      &       6(4)              &         2 (1)              \tabularnewline
N(1710)1/2+                          &   $\Theta_{\pi N}$                &      120-193    &    120(70)              &          155 (21)              \tabularnewline
\rule{0pt}{3ex}
                                    & $\frac{2 |a|_{\eta N}}{\Gamma}$   &        -        &      12(4)      \%      &           14(3)  \%           \tabularnewline
                                    &    $\Theta_{\eta N}$ 	            &        -        &         0(45)           &            21 (7)              \tabularnewline
\hline
	                                &    $M$	                        &  1860(35)       & 1860(35)$^{[18]}$       &   1875 (9)                   \tabularnewline
                                    &    $\Gamma$                       &   250(70)       &    250(70)              &          232  (15)             \tabularnewline
                                    &  $ |a|_{\pi N}$	                &     6(4)        &      6(4)               &           3(1)                \tabularnewline
N(1880)1/2+                          &   $\Theta_{\pi N}$                &     80(65)      &      80(65)             &           107(16)              \tabularnewline
\rule{0pt}{3ex}
                                    &  $\frac{2 |a|_{\eta N}}{\Gamma}$  &        -        &      11(7) \%           &             6(1) \%            \tabularnewline
                                    &   $\Theta_{\eta N}$ 	            &        -        &       -75(55)           &            -131(26)             \tabularnewline
\hline
	                                &   $M$	                            &  2120(40)       &  2100  $^{[19]}$        &      2171(24)                   \tabularnewline
                                    &   $\Gamma$                        &    240(80)      & 500  \hspace*{0.2cm}    &          210(48)                \tabularnewline
                                    &  $ |a|_{\pi N}$	                &   14(7)         &         -               &          15 (5)                \tabularnewline
N(2100)1/2+                         &   $\Theta_{\pi N}$ 	            &    35(25)       &         -               &          -50 (8)               \tabularnewline
\rule{0pt}{3ex}
                                    &  $\frac{2 |a|_{\eta N}}{\Gamma}$  &        -        &         -               &           16 (4) \%              \tabularnewline
                                    &   $\Theta_{\eta N}$ 	            &        -        &         -               &           -139 (19)              \tabularnewline
\hline \hline
\end{tabular}
\end{center}
\end{table}

The results presented in this table confirm that both, pole positions and residua, generally lie within one standard deviation intervals when compared with the published results. As discussed above, better agreement cannot be expected. Surprisingly, the weak and poorly determined N(2100)1/2+ resonance from the Bonn-Gatchina model is, in the MC L+P fit, not only well and confidently reproduced, but also  necessary.

\clearpage

As a conclusion, we state that the generalization of the L+P model to a multi-channel case, as described in this paper, provides a powerful but simple and precise method to extract pole positions from coupled processes (coupled-channel models) and correlated quantities ($E_{l_\pm}$ and  $M_{l_\pm}$ in photo-production), and that this is the first method which can be directly used to extract pole positions from partial-wave amplitudes extracted from experimental data.

\end{document}